\documentstyle[12pt,epsfig]{article}
 
\def\er#1#2{\relax\ifmmode{}^{+#1}_{-#2}\else$^{+#1}_{-#2}$\fi}
\newcommand{\be}{\begin{equation}}
\newcommand{\bea}{\begin{eqnarray}}
\newcommand{\ee}{\end{equation}}
\newcommand{\eea}{\end{eqnarray}}

\begin{document}
\begin{center}
{\Large \bf Parton Distributions Functions of Pion, Kaon and Eta
 pseudoscalar mesons in the NJL model} \\ [2em]

  R. M. Davidson\footnote{e-mail:davidr@rpi.edu}   \\ 
  Department of Physics, Applied Physics and Astronomy, \\ 
  Rensselaer Polytechnic  Institute, Troy, New York,
  12180-3590. USA.  \\ 
  E. Ruiz Arriola\footnote{e-mail:earriola@ugr.es} \\  Departamento de
  F\'{\i}sica Moderna \\ Universidad de Granada, E-18071 Granada. 
  Spain.\\

\end{center} 

%\vskip.2cm
\begin{abstract}
Parton distributions of pseudoscalar $\pi$,$K$ and $\eta$ mesons
obtained within the NJL model using the Pauli-Villars regularization
method are analyzed in terms of LO and NLO evolution, and the valence
sea quark
and gluon parton distributions for the pion are obtained at $Q^2 =
{\rm 4 GeV}^2 $ and compared to existing parametrizations at that
scale. Surprisingly,
the NLO order effects turn out to be small compared
to the LO ones. The valence distributions are in good agreement
with experimental analyses, but
the gluon and sea distributions come out to be
softer in the high-x region and harder in the low-x region than the
experimental analyses suggest.
\vskip.3cm 
\noindent
\centerline{\it 
PACS numbers: 13.60.Hb, 12.39.Ki, 12.39.Fe   \\}

\centerline{\it Keywords: Pseudoscalar mesons, Parton distribution
functions,} \centerline{\it Structure functions, QCD evolution,
Nambu--Jona-Lasinio model}
\end{abstract}

\section{Introduction}

The study of structure functions of hadrons in the Bjorken limit and
high enough $Q^2$, as probed in inclusive deep inelastic
scattering (DIS), is traditionally considered the domain of
perturbative QCD since the running coupling constant $\alpha (Q^2)$,
becomes small \cite{Ro90}. Present day QCD leading order (LO) and next
to leading order (NLO) phenomenological calculations can relate
leading twist contributions to structure functions among different
momentum scales through the well known linear integro-differential
Gribov-Lipatov-Altarelli-Parisi (GLAP) equations \cite{AP77}. This
makes sense if $Q^2$ is high enough so that only leading twist
logarithmic corrections contribute and higher twist power-like
corrections are negligible. To start with, some theoretical or
experimental nonperturbative
profile function is needed as initial condition for the
GLAP equations. In the nucleon case, QCD scaling violations have been
confirmed by relating experimental partonic distributions at many $
Q^2 $ values, and many phenomenological parametrizations have been
proposed \cite{GRV95,MRST98,GRV98,La00}. Naturally, these
parametrizations are under continuous update to incorporate
increasing information obtained from current experiments. The net
result is that uncertainties in the parton distribution functions of
the nucleon include not only a large body of experimental data, but
also theoretical NNLO or higher twist error estimates which provide a,
perhaps inaccurate, but undoubtedly systematic description within a
large region in the $x,Q^2$ plane (see, e.g., the talks in
Ref.~\cite{DIS2001}).

In contrast with the nucleon case, our present knowledge
of parton distribution functions for other hadrons is rather poor.
As suggested long ago \cite{JR80}, it is possible to estimate
distribution functions using
constituent quark models to evaluate the
low energy initial condition under the assumption that the gluon and
sea content of hadrons vanish at the corresponding low energy
resolution scale, and dynamically generate them by QCD evolution to higher
$Q^2$ scales. These estimates can then be used to test the sensitivity
of various experiments to the distributions of interest.
Recently, this approach has also been applied to
generalized parton distributions (GPD's)
\cite{bar82,dit88,mul94,ji97,rad97,col97}, which are generalizations of
the usual parton distributions considered in this work and are related
via a sum rule to the elastic form factors. Unlike the usual parton
distributions, the GPD's are not directly measurable as observables are
always expressed in terms of them via a convolution formula. This has
lead some \cite{pet98} to conclude that, at present, model calculations
of the GPD's at a low scale are needed to assess the sensitivity of
various observables to the GPD's. In practice, this requires an evolution
of the GPD's calculated at the low energy scale to the scale relevant
to the experiment. As $\alpha$ is rather large at the low energy scale,
one must worry about the use of perturbative evolution to connect
the low energy model with the high energy data. Thus, it seems prudent
to test this procedure in a situation where data are available to compare
with theory, i.e., the usual parton distributions. To test the validity
of this approach, it is necessary to compare the LO and NLO results not
only for the valence distributions, but also for the sea and gluon
distributions.

From a theoretical viewpoint, pseudoscalar mesons and specifically
$\pi$ and $K$ mesons are particularly distinguished hadrons since most
of their low energy properties follow the patterns dictated by chiral
symmetry. Actually, we do not expect to understand the properties of
any hadron better than the pion, as Chiral Perturbation Theory
suggests. By extension, one might think that the parton structure of a
pion is the simplest one to consider provided chiral symmetry
constraints, i.e., spontaneous and explicit chiral symmetry breaking,
are properly incorporated. The recent work \cite{AS01} clarifies this
point regarding explicit chiral symmetry breaking; ChPT allows one to
systematically compute chiral corrections to the moments of structure
functions, but says nothing about the soft pion limit. Each moment
corresponds to a undetermined low energy parameter which renormalizes
a local operator. On the other hand, improved QCD sum rules have also
been employed \cite{IO00} to determine the quark distribution
functions of the pion in the intermediate $x$ region, $0.15 < x < 0.7
$ at $Q^2= 2 {\rm GeV}^2 $ where the model is applicable. The absolute
normalization becomes a problem since some ansatz for the
distributions must be made outside this $x-$range.

Among the quark models where
spontaneous breaking of chiral symmetry plays a dominant role, the
Nambu--Jona-Lasinio (NJL) model provides a particular example of a
chiral quark model where a unified picture of vacuum, mesons and
nucleons is achieved \cite{NJL}. Pseudoscalar mesons appear as
quark-antiquark excitations of the spontaneously broken vacuum. Several
calculations of the pion structure functions within
chiral quark loop models have been made and many different
results for the initial conditions have been obtained. One important
and tricky reason for the discrepancies lies in the use of different
regularization procedures. As the bosonized version of the
NJL model is similar to other quark-loop models of
the pion (the
$\pi qq$ coupling is $\gamma_5$-like), we think it of interest to
briefly review them and comment on the main differences.
The use of different regularizations might be regarded as an
objection to the NJL model itself. However, not every regularization
scheme can be considered acceptable. Actually, some of the quark-loop
calculations violate some necessary conditions on the
regularization. We argue in the following that in some cases one
should blame the regularization scheme instead of the model.
At a formal level, the process of going from the
hadronic to the distribution function can be done by using the
so-called quark-target scattering formula \cite{Ja85}. A large body of
quark loop model calculations have been done making use of these ideas
\cite{SS93,FM94,DR95,SW95,JM97,BH99,WRG99,He00,DT00,HRS01,PR01,Ru01}. The
problem of proceeding in that way is that the distribution function
may turn out to be non-normalizable.

Unfortunately, no first principle calculations of structure functions
for pseudoscalar mesons are yet available with the exception of some
standard lattice calculations of the lowest moments
\cite{MS87,MS88,Be97,DP00}, albeit in the quenched approximation and
subjected to well-known problems with chiral extrapolations. In
addition, the reconstruction of the $x$ dependent structure function
via some inverse moments method is strongly biased in the
intermediate and low $x$ regions.  The transverse lattice approach
employed in Refs.~\cite{Da01,BS01} offers the possibility of directly
computing structure functions in $x$-space. In any case, as one
might expect from the quenched approximation, the lattice results
provide a larger momentum fraction of valence quarks than those suggested by
phenomenological analyses~\cite{SMRS92,GRS98,GRS99}.

Although not as well determined as the nucleon, the parton structure
of the pion has been analyzed on a phenomenological level
\cite{SMRS92} and a simple parametrization at $Q^2 = 4 {\rm GeV}^2$
has been given. The valence quark distributions extracted in this work
\cite{SMRS92} from Drell-Yan experiments \cite{Co89} seem well
determined, whereas the gluon distribution as obtained from the
analysis of prompt-photon emission data \cite{Au89} is less well
determined.  On a phenomenological level, the constituent model
proposed in Ref.~\cite{AP96} for the valence distributions of the pion
has been further extended to the sea and gluon distributions
\cite{GRS98} and the $K/\pi$ valence up-quark ratio. In these
calculations and in the recent update \cite{GRS99} the required total
valence momentum fraction in the pion at $Q^2 = 4 {\rm GeV}^2 $ is
taken to be the same as for the nucleon, $\langle x V_\pi \rangle =
\langle x V_N \rangle = 0.40 $, a bit below the value $ \langle x
V_\pi \rangle = 0.47 $ of Ref.~\cite{SMRS92}. As different data sets
have been fitted and different nucleon parton distributions have been
used in the different analyses, it is not clear what to make of the
differences. In addition, although Ref.~\cite{SMRS92} includes error
estimates, the model analysis of Refs.~\cite{GRS98,GRS99} does not
include them, and therefore it is not possible to know if the
differences are significant. Let us note that the E615
experiment~\cite{Co89} suggests the valence density of the pion may be
enhanced by about $20\%$ compared to the proton, and a recent analysis
\cite{kla01} of the ZEUS di-jet data seem to favor the gluon
distributions of Ref.~\cite{SMRS92}. Thus, in determining the
low-energy scale of our model, we use the valence momentum
fraction found in Ref.~\cite{SMRS92}. Finally, we also compare with
the $K^- /\pi^-$ structure function ratio at $Q^2 = 20 {\rm GeV}^2 $,
which has been measured using the Drell-Yan process \cite{Ba80}.

\section{Remarks on Pion parton distribution functions in quark loop models}

In a previous work \cite{DR95}, we found the structure function of the
pion to be a constant function of $x$ in the NJL model in the chiral
limit and in the leading order of a large $N_c$ expansion. To get this
result the use of a suitable regularization method was needed. A
thorough study of regularization methods in the NJL model may be found
in Ref.~\cite{Do92} and we refer to that work for a more detailed
description. By suitable we mean several desirable properties that
should be incorporated, namely:

\begin{itemize} 
\item The connection between the forward Compton amplitude and the
quark-target scattering amplitude is valid only for gauge invariant,
finite amplitudes. For this reason, some gauge invariant
regularization must be imposed on the Compton amplitude. Naive sharp
cut-offs are not acceptable from this viewpoint. In addition, this way
of proceeding represents a further advantage, since in the NJL model
it is only known how to regularize closed quark loops. The quark
target scattering amplitude corresponds to an open quark line.
\item The regularization must produce exact scaling in the Bjorken
limit. The main reason is that this is the only way we know how to
extract the leading, and eventually higher, twist contributions for which
QCD evolution is known. This eliminates proper-time regularization,
since it produces unrealistic scaling violations.  
\item The regularization must also be able to work away from the
chiral limit, but without spoiling the QCD anomaly. The former
condition precludes a single Pauli-Villars subtraction.  
\item The regularization should allow calculations in both Minkowski
and Euclidean space, i.e., dispersion relations must be fulfilled. This
turns out to be very convenient for DIS calculations, since cutting
rules may be used.
\item The resulting distributions should satisfy the normalization
condition and the momentum sum rule.

\end{itemize} 

We found in Ref.~\cite{DR95} that the Pauli-Villars with two
subtractions fulfills the desired requirements. In addition, the
Pauli-Villars scheme does not spoil the good description of other low
energy hadronic properties found in the NJL model
\cite{NJL,Ru91,SRG92}, fulfills dispersion relations \cite{DR96}, and
allows one to regularize the Dirac sea of the chiral soliton away from the
chiral limit \cite{WRG99}. Taking $\pi^+$ for definiteness, one gets
in the chiral limit
\begin{eqnarray}
u_{\pi^+ } (x,Q_0^2 ) = \bar d_{\pi^+} (x, Q_0^2 ) = \theta(x)
\theta(1-x) \; .
\label{result} 
\end{eqnarray} 
The results for $m_\pi \neq 0 $ are displayed for completeness in the
Appendix.  By construction, Eq.(\ref{result}) is consistent with
chiral symmetry. The result was obtained by several means within the
NJL model either using Pauli-Villars regularization \cite{DR95,WRG99}
on the virtual Compton amplitude or imposing a transverse cut-off
\cite{BH99} upon the quark-target amplitude. This result has been
recently re-derived \cite{Ru01} in a chiral quark model solving
chiral Ward identities by using the so-called gauge technique
\cite{DW77}. The easiest way to understand Eq.~(\ref{result}) is
perhaps in terms of phase space arguments and point couplings (i.e.,
constant matrix elements) \cite{Ru98b}. For a massless pion this is
justified since intermediate states in the quark-target amplitude
have $p_n^+ = m_\pi ( 1-x) \to 0 $ and the low momentum components of
$\pi \bar q q $ matrix element dominate.  Let us mention that
Eq.(\ref{result}) disagrees with other NJL calculations, due to the
use of different regularizations. If the virtual Compton amplitude is
used with a four-dimensional cut-off \cite{SS93} or the quark-target
amplitude is used with Lepage-Brodsky regularization \cite{BH99},
different shapes for the quark distributions are obtained. The
null-plane \cite{FM94} NJL model with sharp cut-off \cite{SS93},
light-cone (LC) quantized NJL model \cite{He00} and spectator model
\cite{JM97} calculations also produce different results. In all cases,
the use of momentum dependent form factors or non-gauge invariant
regularizations make the connection between Compton amplitude and
quark-target amplitude doubtful and, furthermore, spoil
normalization. The results based on a quark loop with momentum
dependent quark masses \cite{DT00,HRS01,PR01} seem to produce a
non-constant distribution. Recent calculations on the transverse
lattice reveal \cite{Da01} either an almost flat structure very much
resembling Eq.~(\ref{result}) at a scale $Q^2 = 1 {\rm GeV}^2 $ or a
more bumped form \cite{BS01}. The reason for the discrepancy between
these two transverse-lattice calculations is not obvious to us.

In this paper we study within LO and NLO the parton content of
pseudoscalar mesons, namely $\pi$, $K$ and $\eta$ including valence,
gluon and sea distributions, thus extending our previous work
\cite{DR95} where only the initial conditions were presented and the
LO evolution for the valence distributions. There, we analyzed the LO
valence contribution and impressive agreement with SMRS \cite{SMRS92}
parametrization at $Q^2 = 4 {\rm GeV}^2 $ was obtained. Encouraged by
this success we extend our analysis to the sea and gluon distributions
both in LO and NLO evolution.

\begin{figure}[t]
\begin{center}
\epsfig{figure=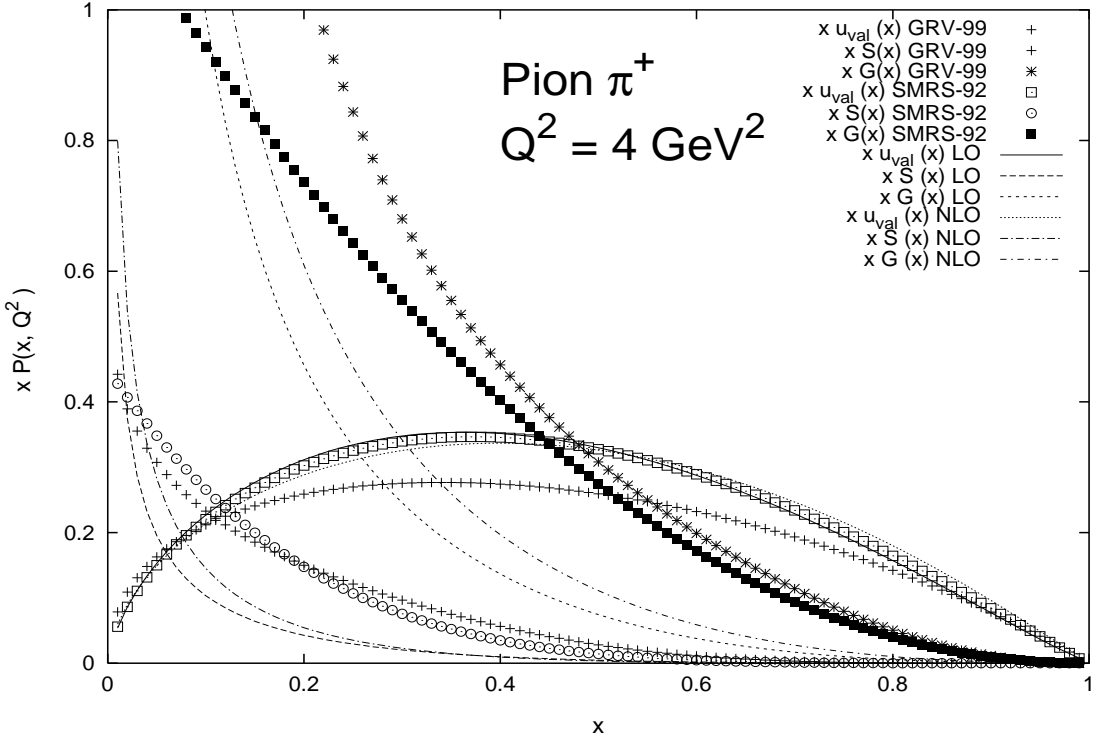,height=10cm,width=15cm}
\end{center} 
\caption{Valence, gluon and sea distributions in the pion, $\pi^+$, at
\protect{$Q^2 = 4 {\rm GeV}^2 $} in the NJL model compared with
phenomenological analysis for the pion SMRS92 \cite{SMRS92} and GRS99
\cite{GRS99}. We take the valence momentum fraction $ \langle x V
\rangle_\pi = 0.47 $ at $ Q^2 = 4 {\rm GeV}^2 $. }
\label{fig:pion}
\end{figure}

\begin{figure}[t]
\begin{center}
\epsfig{figure=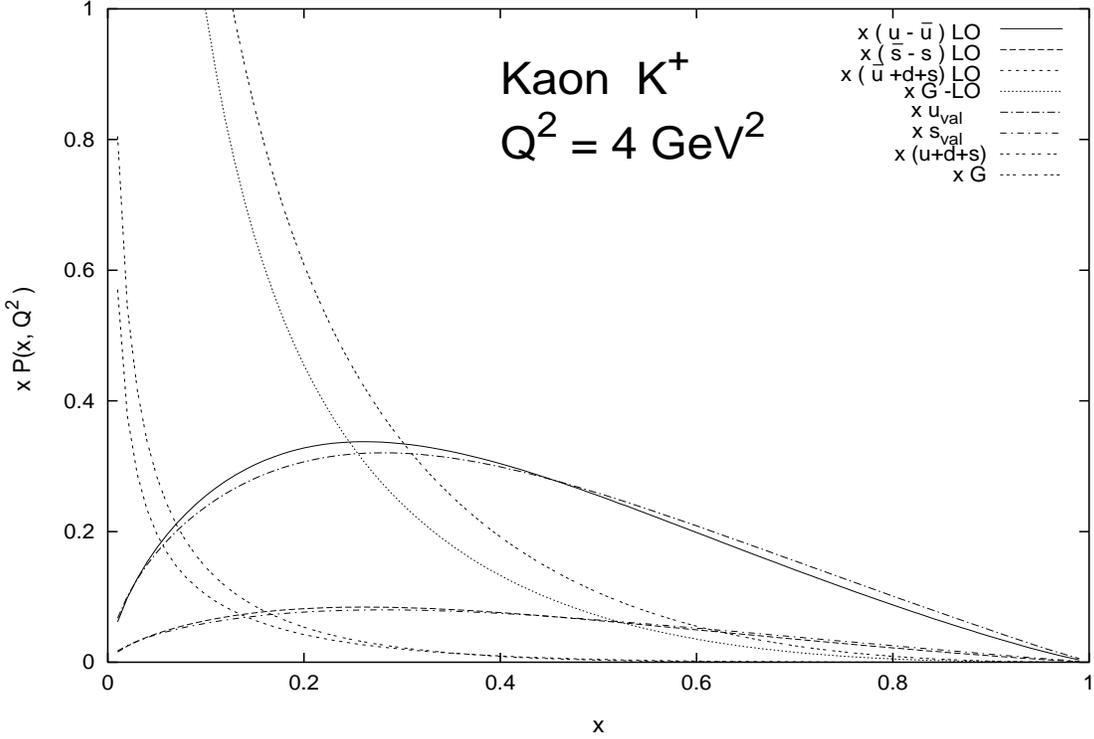,height=10cm,width=15cm}
\end{center} 
\caption{Valence, gluon and sea distributions in the kaon, $K^+$, at
\protect{$Q^2 = 4 {\rm GeV}^2 $} in the NJL model. We take the total
valence momentum fraction $ \langle x V \rangle_\pi = 0.47 $ at $ Q^2
= 4 {\rm GeV}^2 $.}
\label{fig:kaon}
\end{figure}

\begin{figure}[t]
\begin{center}
\epsfig{figure=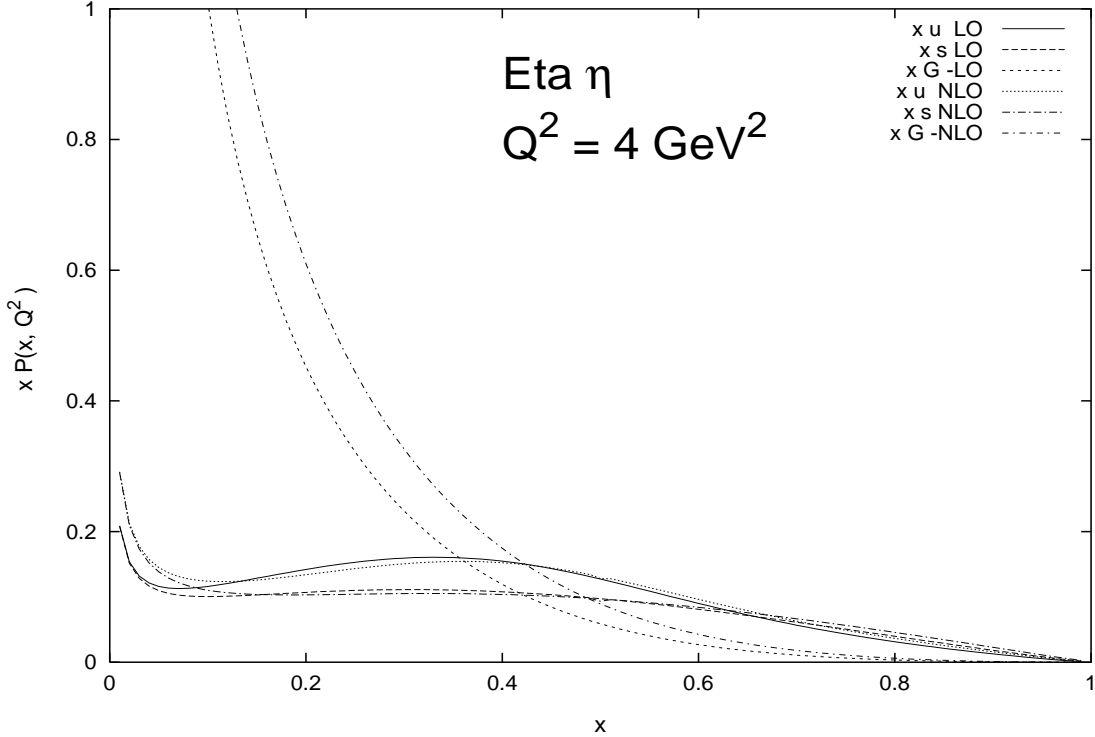,height=10cm,width=15cm}
\end{center} 
\caption{up and strange quark and gluon LO and NLO distribution
functions in the $\eta$-meson, at \protect{$Q^2 = 4 {\rm GeV}^2 $} in
the NJL model.}
\label{fig:eta}
\end{figure}

\begin{figure}[t]
\begin{center}
\epsfig{figure=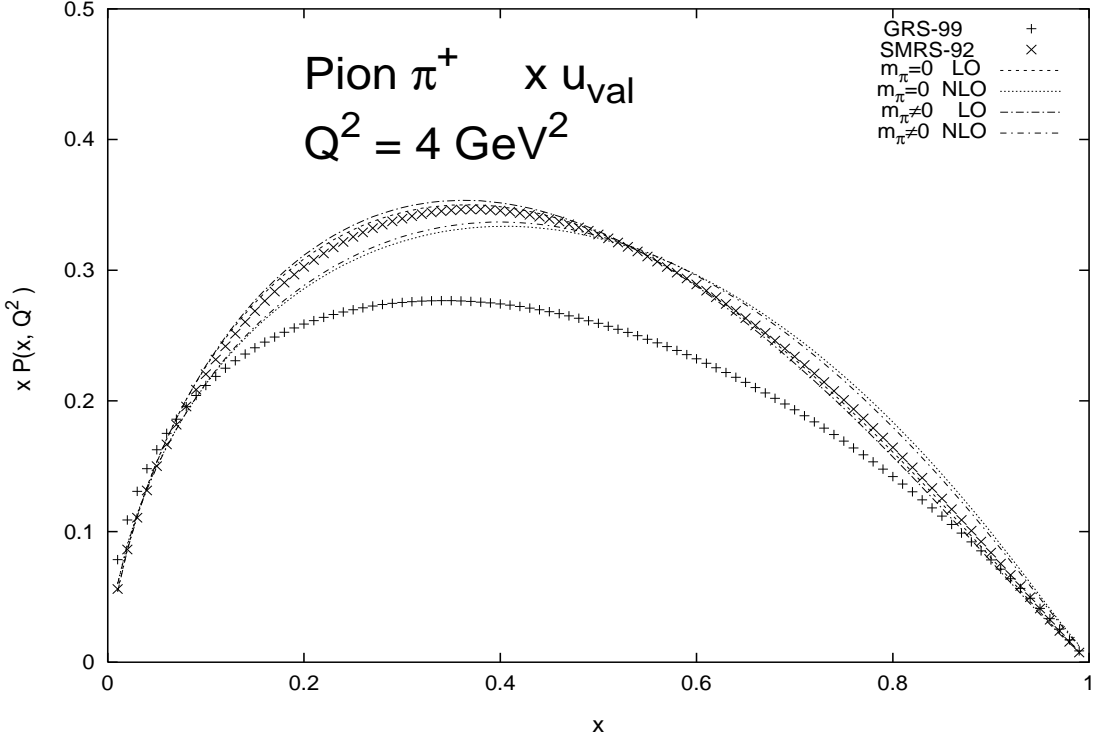,height=10cm,width=15cm}
\end{center} 
\caption{Chiral corrections to the u-quark valence LO an NLO
distribution functions at \protect{$Q^2 = 4 {\rm GeV}^2 $} compared
with phenomenological analysis for the pion SMRS92 \cite{SMRS92} and
GRS99 \cite{GRS99}. As suggested in Ref.~\cite{SMRS92} we take $
\langle x V \rangle_\pi = 0.47 $ at $ Q^2 = 4 {\rm GeV}^2 $.}
\label{fig:chiral}
\end{figure}

\begin{figure}[t]
\begin{center}
\epsfig{figure=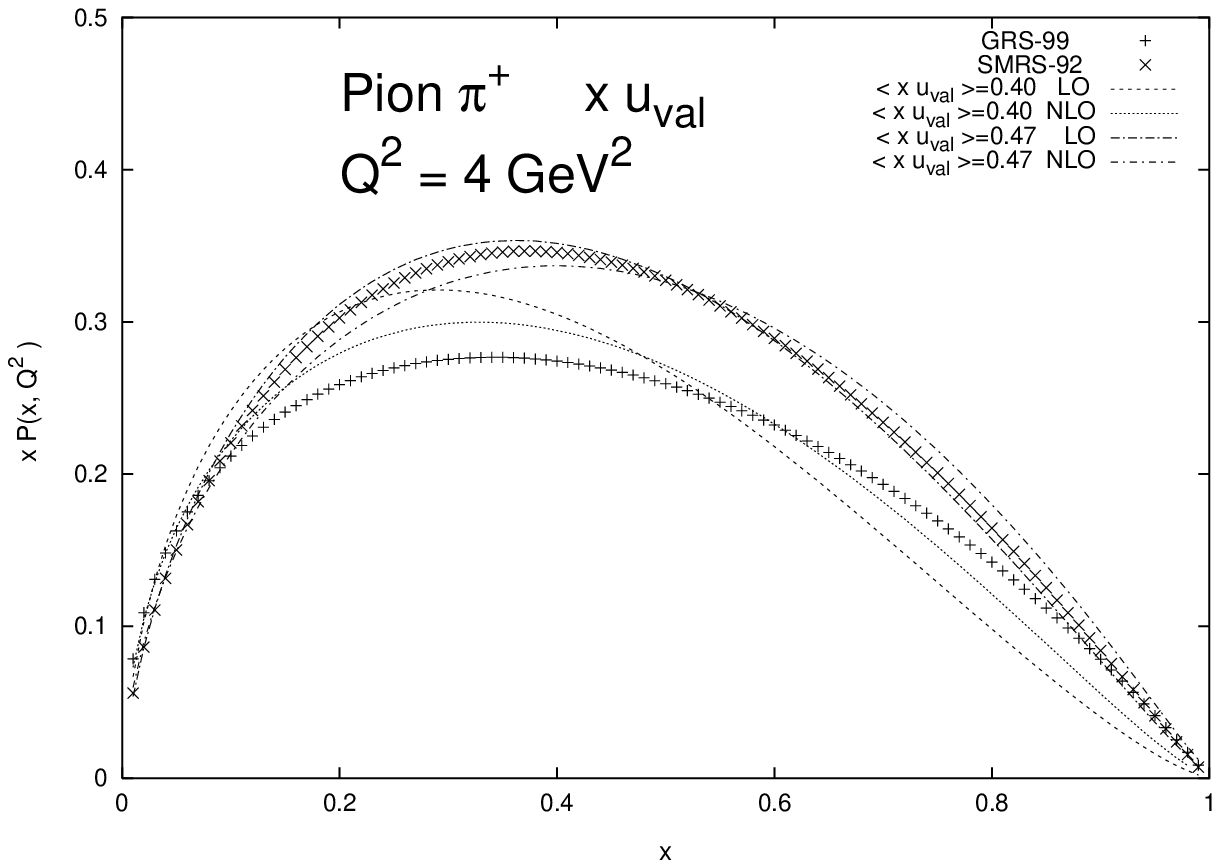,height=10cm,width=15cm}
\end{center} 
\caption{Dependence of the u-quark valence LO an NLO distribution
functions at \protect{$Q^2 = 4 {\rm GeV}^2 $} on the momentum fraction
at that scale, compared with phenomenological analysis for the pion
SMRS92 \cite{SMRS92} which takes $ \langle x V \rangle_\pi = 0.47 $
and GRS99 \cite{GRS99} where $ \langle x V \rangle_\pi =0.40$ is
used.}
\label{fig:4047}
\end{figure}

\section{Numerical Results} 

\subsection{Momentum fraction analysis}

To perform the evolution, one must determine the scale $Q_0$ of
the model. We determine this scale by fitting the valence
quark momentum fraction at 4 GeV$^2$.
For definiteness, we take the running strong coupling constant at the
$Z$ mass, $M_Z =91.12 {\rm GeV} $, to be $\alpha (M_Z^2) =0.116 $ and
evolve it down by exactly solving the differential equation
\begin{eqnarray}
{d \alpha \over dt} = \beta ( \alpha ) = - \alpha \left[ \beta_0
\left({\alpha \over 4\pi}\right) + \beta_1 \left({\alpha \over 4\pi}
\right)^2 + \cdots \right] 
\label{eq:beta}
\end{eqnarray} 
where $t= {\rm ln} (Q^2 / Q_0^2 ) $ and $\alpha=g^2 / (4\pi) $.
We take the number of active flavors to diminish by one
unit each time a quark threshold is crossed, i.e., $N_F (Q^2) =
\sum_{i=u,ds,c,b,t} \theta ( Q^2 - m_i^2 ) $, with $m_b=4.5 {\rm GeV}
$ and $m_c = 2.0 {\rm GeV}$. For $N_F=3 $ one has
$\beta_0=9$, $\beta_1 = 64$.  This yields the value $\alpha ( 4 {\rm
GeV}^2 ) = 0.284 $.
Below that scale we fix the number of flavors equal
to three, since we consider evolution below charm threshold. 
The previous formula, Eq.~(\ref{eq:beta}), is used to transform the
variable $t$ into the variable $\alpha$, by exactly\footnote{By `exact' we
mean
solving the differential equation with a numerical accuracy much greater
than the experimental uncertainty on $\alpha$.}
solving the differential equation. Since we numerically perform
the NLO evolution of the sea and gluon distributions, it is convenient
to specify the initial $\alpha_i$
at $t_i$ and numerically integrate Eq. (\ref{eq:beta}) to $t_f$ to
obtain $\alpha_f$. We note, however, that it is also possible to find an
implicit solution for $\alpha_f$ in terms of $\alpha_i$, $t_i$ and
$t_f$. Specifically, we find
\begin{eqnarray}
{ 1 \over \alpha_f} + { \beta_1 \over 4\pi \beta_0} \ln
\left( { \alpha_f \over \alpha_i} \right) - { \beta_1 \over 4\pi \beta_0} \ln
\left( { 1+{\beta_1 \over 4\pi\beta_0}\alpha_f \over
1+{\beta_1 \over 4\pi\beta_0}\alpha_i} \right) &=& { \beta_0 \over 4\pi}
(t_f -t_i) + {1 \over \alpha_i} \nonumber \\
 &=& { \beta_0 \over 4\pi} \ln \left( { Q_f^2 \over Q_i^2} \right ) 
+ {1 \over \alpha_i} \; .
\end{eqnarray}
Although we are not aware of an analytic solution for $\alpha_f$ in terms
of the other parameters, this equation may be solved numerically very
quickly and accurately using Newton's method. For example, taking
$Q_i = m_c$ and $\alpha_i = 0.284$, and using this $\alpha_i$ as the
initial seed for $\alpha_f$, one obtains at least eight-significant
digit accuracy for $\alpha_f$ after at most ten iterations all the
way down to $Q_f$ of 0.4 GeV. This form also enables one to determine
at what scale $\alpha_f$ diverges. For $\alpha_f \rightarrow \infty$,
we obtain
\begin{equation}
{2\pi \beta_0 \over \beta_1 \alpha_f^2} \approx { 1 \over \alpha_i }+ 
{ \beta_0 \over 2\pi} \ln \left( { Q_f \over Q_i} \right) -
{ \beta_1 \over 4\pi \beta_0} \ln \left( 1+ { 4\pi \beta_0 \over \beta_1
\alpha_i } \right) \; .
\end{equation}
Evidently, $\alpha_f$ diverges when the right-hand-side of the above
equation vanishes, which happens at a scale of $Q_f \approx$ 0.365 GeV.

The non-singlet momentum fraction
satisfies the differential equation
\begin{eqnarray} 
\beta(\alpha) { d V_2 (\alpha ) \over d\alpha } = \gamma_{2,NS}
(\alpha) V_2 (\alpha ) 
\label{eq:v2}
\end{eqnarray} 
where $V_2 (\alpha) $ can be any non-singlet quark distribution.
Up to two loops one obtains the expansion \cite{AP77} 
\begin{eqnarray} 
\gamma_{2,{\rm NS}} (\alpha) =
\gamma_{2,{\rm NS}}^{(0)} \left({\alpha \over 4\pi}\right)
+ \gamma_{2,{\rm NS}}^{(1)} \left({\alpha \over 4\pi}\right)^2
+ \cdots 
\end{eqnarray} 
To proceed further, we use the results from Ref.~\cite{SMRS92} where
it was found that at $Q^2 = 4 {\rm GeV}^2 $ valence quarks carry $47
\%$ of the total momentum fraction in the pion, e.g., for $\pi^+$,
\begin{eqnarray} 
\langle x \left( u_\pi - \bar u_\pi + \bar d_\pi - d_\pi \right) \rangle =
0.47 \qquad {\rm at} \qquad Q^2 = 4 {\rm GeV}^2 \; .
\end{eqnarray} 
Evolving downwards, we
get that for $\alpha_0 = 1.89 (1.487) $ valence quarks carry $100\%$
of the total momentum in the pion in LO (NLO).

\begin{figure}[t]
\begin{center}
\epsfig{figure=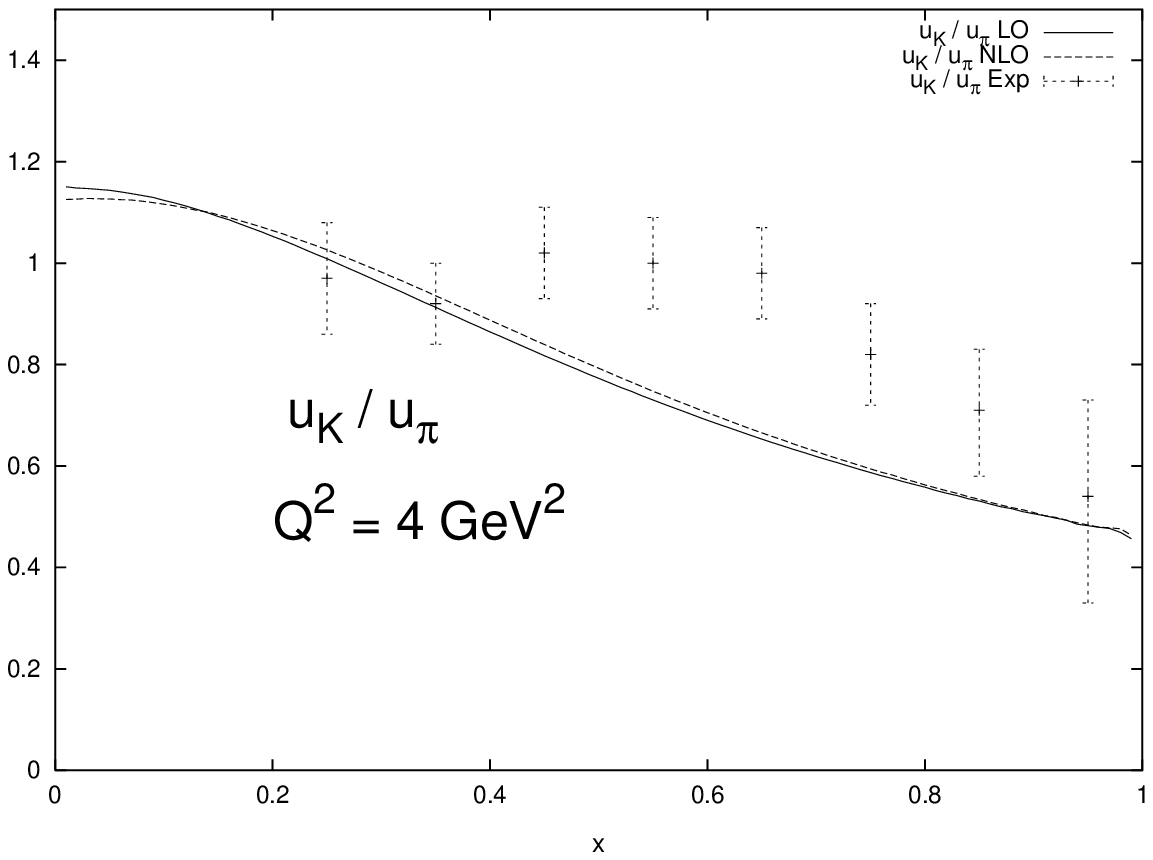,height=10cm,width=15cm}
\end{center} 
\caption{Valence u-quark kaon/pion ratio of LO and NLO distribution
functions in the NJL model at \protect{$Q^2 = 4 {\rm GeV}^2 $}
compared with phenomenological analysis. We take a total valence
momentum fraction $ \langle x V \rangle_\pi = \langle x V \rangle_K =
0.47 $ at $ Q^2 = 4 {\rm GeV}^2 $. Experimental data from
Ref.~\cite{Ba80}.}
\label{fig:uka:upi}
\end{figure}

\subsection{Pion structure functions} 

Having determined $Q_0$ of our model, we evolve the structure functions
to $Q^2 = 4$ GeV$^2$ using
the scheme presented in Ref.~\cite{Ru98a}, which
requires an analytical formula for the moments of
the distribution function. In the chiral limit, $m_\pi=0$, the moments
may be trivially computed. Away from it, $m_\pi \neq 0$ , they can be
expressed in terms of hypergeometric functions $_2 F_1$, but
it is more convenient, and just as accurate, to make a polynomial
approximation in $x-$space and then compute analytically the
moments.  For completeness, the result of such a fit for both $u(x) $
and $\bar d(x)$ is presented in the appendix. For the $\eta$, it is more
convenient to make an expansion in terms of $x(1-x)$, as is discussed
in the appendix. In this work, we take $m_{\pi}$= 139.6 MeV, $m_k =$
494 MeV, $f_{\pi}=$ 93.3 MeV and $M_u =M_d =$ 280 MeV, resulting in
$M_s$= 527 MeV, $\Lambda=$ 870 MeV and $m_{\eta}=$ 501 MeV (exp.~549 MeV).

Our LO and NLO valence, sea and gluon distribution functions evolved
from the quark model point, $Q_0^2$, where the valence quarks carry all the
momentum, to the point $Q^2 = 4 {\rm GeV}^2$ where gluon and sea
distributions are dynamically generated, are shown in
Fig.~(\ref{fig:pion}). They are compared to the phenomenological
analysis of Refs.~\cite{SMRS92} and \cite{GRS99}. The remaining distributions 
trivially fulfill 
\begin{eqnarray} 
\bar d_{\pi^+}(x,Q^2) = u_{\pi^+}(x,Q^2) \qquad \bar u_{\pi^+}(x,Q^2)
= d_{\pi^+}(x,Q^2) \qquad \bar s_{\pi^+}(x,Q^2) = s_{\pi^+}(x,Q^2)
\end{eqnarray} 
as a consequence of our initial condition and properties of evolution.
The LO valence result was already presented in our previous work
\cite{DR95}. We see here that NLO evolution does not make a big
difference, providing some confidence in perturbative evolution,
even though the quark model point corresponds to $\alpha$'s larger
than unity. Actually, it has been suggested that the natural expansion
parameter for DIS is $\alpha / \pi$, which in our case is about a half,
$\alpha (Q_0) /\pi \sim 0.5$.

As can be seen in Fig.~(\ref{fig:chiral}), the finite pion mass effects
turn out to be rather small because chiral corrections to the initial
condition are small within the model at the one loop level. While it is
conceivable that pion loop effects could provide, as is frequently the
case, some logarithmic enhancement to chiral corrections,
it is a feature of GLAP evolution equations that upward evolution
tends to wash out the differences in the initial condition.

We finish our discussion on the pion parton distribution by comparing
in Fig.~(\ref{fig:4047}) the results obtained by taking either $
\langle x V \rangle_\pi = 0.47 $ as suggested by the SMRS92 analysis
\cite{SMRS92} or $ \langle x V \rangle_\pi =0.40$ as implied by the
GRS99 parametrization \cite{GRS99}. The sea and gluon distributions
are not shown because their dependence on the momentum fraction is
rather small. As can be deduced from the figure, the shape of the
valence distribution is much better described if, as determined in
Ref.~\cite{SMRS92}, the valence quarks carry $47\%$ of the total pion
momentum at $Q^2 = 4 {\rm GeV}^2 $. Note that, as one
might expect, Fig.~\ref{fig:4047} also illustrates the fact that
reproducing the momentum fraction is not sufficient to accurately
determine the full shape of the distribution functions. From this
point of view the agreement of the NJL evolved valence quark
distribution with the SMRS92 parametrization \cite{SMRS92} is not
entirely trivial. 

For comparison, let us also mention that early lattice calculations of
Ref.~\cite{MS87,MS88} provided $\langle x V_\pi \rangle = 0.64 \pm
0.10$ scale $Q^2 \approx 4.84 \pm 2.2 {\rm GeV}^2$. A recent and more
accurate lattice QCD calculation \cite{Be97} extrapolated to the
chiral limit yields the number $ \langle x V_\pi \rangle = 0.56 \pm
0.02$ at the scale $Q^2 \approx 5.8 {\rm GeV^2}$, a larger value than
suggested by phenomenology \cite{SMRS92,GRS99} and expected from a
quenched approximation. The transverse lattice calculation of
Ref.~\cite{Da01} gives $ \langle x V_\pi \rangle = 0.86 \pm 0.02 $ at
$Q^2 \approx 1 {\rm GeV}^2 $, whereas that of Ref.~\cite{BS01}
provides, still at very low scales $Q^2 \approx 0.4 {\rm GeV}^2 $, a
form for the distribution amplitude surprisingly close to the
asymptotic value, $  6 x (1-x) $. From their parton distribution
function one gets $ \langle x V_\pi \rangle \approx 0.76 $.

\begin{figure}[t]
\begin{center}
\epsfig{figure=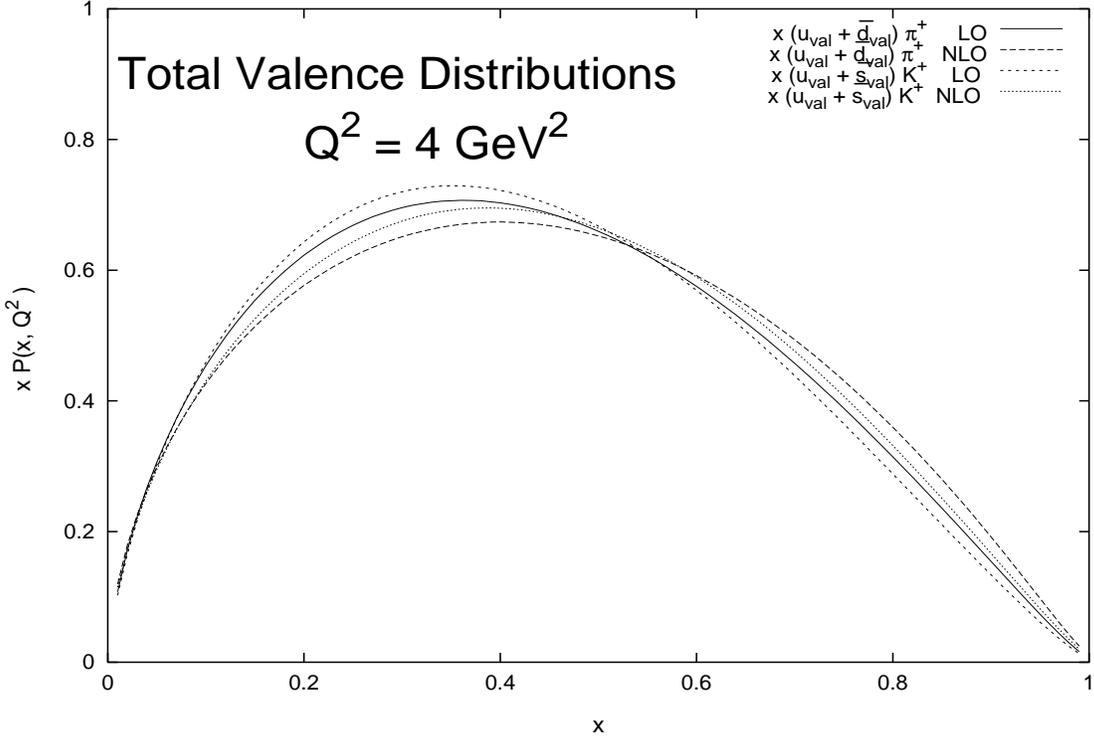,height=10cm,width=15cm}
\end{center} 
\caption{Total valence $\pi^+$ and $K^+ $ LO and NLO distribution
functions in the NJL model at \protect{$Q^2 = 4 {\rm GeV}^2 $}. We
take $ \langle x V \rangle_\pi = \langle x V \rangle_K = 0.47 $ at $
Q^2 = 4 {\rm GeV}^2 $. For $\pi^+$ we define $V=u - \bar u + \bar d -
d$ and for $K^+$ we have $V= u - \bar u + \bar s - s$.}
\label{fig:valence}
\end{figure}

\subsection{Kaon and Eta structure functions} 

For the kaon and eta, we assume the same $Q_0$ as for the pion.
For the $K^+$, this immediately leads to
\begin{eqnarray}
\langle x \left( u_K - \bar u_K + \bar s_K - s_K \right) \rangle =
0.47 \qquad {\rm at} \qquad Q^2 = 4 {\rm GeV}^2 \; .
\end{eqnarray}  
Our LO and NLO evolved results for the $K^+$ parton distributions are
shown in Fig.~(\ref{fig:kaon}). A practical parametrization of the
corresponding initial condition may be found in the Appendix.
Similar to the pion case, there are only
small differences between LO and NLO
evolution. The only known
information regarding $K$ structure functions is the ratio
between the valence up quark distribution in the kaon and the pion, which
was originally reported in Ref.~\cite{Ba80} and has been reanalyzed in
Ref.~\cite{GRS98}. In Fig.~(\ref{fig:uka:upi}) we show the NJL
results, together with the data obtained from
Ref.~\cite{Ba80}. Besides the LO result, already shown in our
previous work~\cite{DR95}, we provide the NLO ratio, which does not
differ much from the former and is in fair agreement with
the experimental data.
For the $K^+$
meson the momentum fraction for the up and strange valence quarks turn
out to be
\begin{eqnarray}
\langle x \left( u_K - \bar u_K \right) \rangle = 0.20
\qquad \langle x \left( \bar s_K - s_K \right) \rangle = 0.27 
 \qquad {\rm at} \qquad Q^2 = 4 {\rm GeV}^2 \; .
\end{eqnarray} 
As could be anticipated from Fig.~(\ref{fig:kaon}), the difference
for these momentum fractions between LO and NLO evolution are small
and do not show up within the presented accuracy. 

Although a phenomenological analysis of the
$\eta$ partonic distributions seems unlikely,
for the sake of completeness we show in
Fig.~(\ref{fig:eta}) our results for the $\eta$ meson. We do this by
evolving from the scale where $\alpha=1.89 (1.49)$ at LO (NLO) to $
Q^2 = 4 {\rm GeV}^2$ the NJL distributions conveniently parametrized
in the Appendix. As explained in our previous work \cite{DR95},
our description relies
on a very particular ansatz which provides flavor mixing without
quark mass mixing.  For the
momentum fractions, we obtain
\begin{eqnarray}
\langle x  u_\eta  \rangle = \langle x  d_\eta  \rangle = 0.10 
\qquad \langle x  s_\eta  \rangle = 0.08 
\qquad {\rm at} \qquad Q^2 = 4 {\rm GeV}^2 \; .
\end{eqnarray}

As we have noted, the differences in parton distribution functions for
massless and massive pions are tiny. In fact, even for $K$ and $\eta$,
many of the distributions are close to those of the massless pion.
By comparing Fig.~(\ref{fig:pion}),
Fig.~(\ref{fig:kaon}) and Fig.~(\ref{fig:eta}) we point out the strong
similarities in the gluon parton distributions between the $\pi$,$K$
and $\eta$ mesons. Likewise, we also find very similar shapes for the
sea distributions in the $\pi$ and $K$ mesons, see
Fig.~(\ref{fig:pion}) and Fig.~(\ref{fig:kaon}), as well as in the
total valence distributions, see Fig.~(\ref{fig:valence}). This is in
agreement with having identical total valence momentum fractions for
the pion and the kaon at $Q^2= 4 {\rm GeV}^2 $.

\section{Conclusions}

In the present work, we have computed the parton distribution
functions of the lowest pseudoscalar mesons, namely $\pi$, $K$ and
$\eta$. To this end we have used the Nambu--Jona-Lasinio distribution
functions at the low resolution scale found in our previous work. In
common with state of the art calculations, the corresponding sea and
gluon distribution functions vanish at that scale, and are dynamically
generated through standard GLAP evolution to higher $Q^2$-values at
LO and NLO approximations. For both $\pi$ and $K$ we have assumed that the
valence quarks carry $47\%$ of the total momentum fraction at 4 GeV$^2$.
Despite the fact that $\alpha (Q_0 )/\pi \approx 0.5$,
the differences between LO and NLO evolution are small.
The agreement between the u-quark valence distribution
in the pion in the NJL model at LO and the phenomenological analyses
is not spoiled at NLO.
In addition, we have confirmed at NLO the successful
description at LO of the ratio of the valence up quark content in the
kaon with respect that of the pion.
This provides
one with some confidence in the validity of this approach to the study
of structure functions, or GPD's in general.

We have also presented LO and NLO sea
and gluon distributions of the pseudoscalar mesons. For the pion, we
find disagreement with the phenomenological expectations; the gluon
and sea distributions come out to be softer in the high-x region and
harder in the low-x region than the experimental analysis suggests. We
have also provided results for the $\eta$ meson, which interest seems
only theoretical, given the lack of experimental data. Our analysis,
however, reveals some clear trends: all gluon distributions look
strikingly similar, and the total valence $\pi$ and $K$ distributions do
not differ much. We hope these observations to be useful to get
further insight and guidance into the theoretical description of the
poorly known meson structure functions.

\section*{Appendix} 

The $\pi$, $K$ and $\eta$ structure functions found in
Ref.~\cite{DR95} may be conveniently written as
\begin{eqnarray}
u_\pi (x) = \bar d_\pi (1-x) &=& 4 N_c g_{\pi uu}^2 {d \over d m_\pi^2 }
\left[ m_\pi^2 F_{uu} ( m_\pi^2 , x ) \right] \\ u_K
(x) = \bar s_K (1-x) &=& 4 N_c g_{\pi us}^2 {d \over d m_K^2 } \left[
m_K^2 F_{us} ( m_K^2 , x ) \right] , \\ u_\eta (x) =
\bar u_\eta (x) = d_\eta (x) = \bar d_\eta (x) &=& 4 N_c \left(
{1\over g_{\eta uu}^2} + {2\over g_{\eta ss}^2} \right)^{-1} {d \over
d m_\eta^2 } \left[ m_\eta^2 F_{uu} ( m_\eta^2 , x ) \right]
\\  s_\eta (x) = \bar s_\eta (x) &=& 8 N_c \left( {1\over g_{\eta
uu}^2} + {2\over g_{\eta ss}^2 }\right)^{-1} {d \over d m_\eta^2 } \left[
m_\eta^2 F_{ss} ( m_\eta^2 , x ) \right] 
\label{eq:dist} 
\end{eqnarray}  
in the interval $ 0 < x < 1 $. The Pauli-Villars regularized one-loop
integrals are defined,
\begin{eqnarray}\label{fs}
F_{\alpha\beta} (p^2, x) &=& -{1\over 16 \pi^2} \sum_i c_i \log\left[
-x(1-x)p^2 + (1-x) M_\alpha^2 + x M_\beta^2 +\Lambda_i^2 \right]
\end{eqnarray} 
where $ \sum_i c_i f( \Lambda_i^2 ) = f(0) - f(\Lambda^2) + \Lambda^2
f' (\Lambda^2 ) $.   All other distribution functions are exactly zero, since we
do not have gluons or sea quarks in the model. The meson-quark-quark
couplings are defined in terms of the residues of the poles
in the $q-\bar{q}$ scattering amplitude, and have the precise form
needed to ensure the normalization conditions
\begin{eqnarray}
\langle u_{\pi} (x)\rangle = \langle \bar{d}_{\pi} (x)\rangle &=& 1 \nonumber \\
\langle u_K (x)\rangle = \langle \bar{s}_K (x)\rangle &=& 1 \nonumber \\
\langle u_{\eta} + d_{\eta} +s_{\eta}\rangle &=& 1 \nonumber \\
\langle \bar{u}_{\eta} + \bar{d}_{\eta} +\bar{s}_{\eta}\rangle &=& 1 \; .
\end{eqnarray}
The function $F_{\alpha\beta}$ satisfies the
symmetry relation $F_{\alpha\beta} (p^2 , x) = F_{\beta\alpha} (p^2 ,
1-x) $. This feature, along with the normalization condition, ensures
the momentum sum rule. For the kaon, for example, one obtains
\begin{eqnarray}
\langle x u_K (x)+x\bar{s}_K (x)\rangle &=& \langle x u_K (x)+x u_K
(1-x)\rangle \nonumber \\ = \langle x u_K (x)+(1-x) u_K (x)\rangle &=&
\langle u_K (x)\rangle = 1 \; .
\end{eqnarray}

To apply the evolution method employed in Ref.~\cite{Ru98a} some
analytical formula for the moments is needed. To obtain an approximate
analytic formula for the moments, we note that for $0<x<1$, a convergent
Taylor expansion of $x$ dependence of Eq. (\ref{fs}) exists. Thus, the
distribution functions may be accurately approximated by an $nth$ degree
polynomial, and the accuracy may be increased by keeping higher order
terms. The pion and kaon distribution functions at $Q_0^2$ are
accurately represented by
\begin{eqnarray}
u_{\pi^+} (x,Q_0^2 ) = \bar d_{\pi^+} (x,Q_0^2) &=& 0.9535 + 0.2664 x
  - 0.2074 x^2 \nonumber \\ && -0.1046 x^3 + 0.0190 x^4 +0.0400 x^5 -
  0.0133 x^6 \\ u_{K^+} (x, Q_0^2) &=& 1.1039 + 1.8071 x - 1.0739 x^2
  \nonumber \\ && -16.2227 x^3 +33.1781 x^4 -25.5372 x^5 + 7.1872 x^6
  \\ \bar s_{K^+} (x , Q_0^2 ) &=& 0.4425 + 0.8593 x +1.7623 x^2
  \nonumber \\ && -4.8611 x^3 +13.2997 x^4 -17.5858 x^5 + 7.1872 x^6 \; .
\end{eqnarray} 
In each case, the remaining quark and gluon distribution functions are
assumed to be zero. For the $\eta$ meson, the expansion does not converge
rapidly because one of the expansion parameters is
\begin{equation}
{M^2_{\eta} x(1-x) \over M_u^2 } \le {M^2_{\eta} \over 4M_u^2}
\approx 0.8 \; .
\end{equation}
The convergent series for $u_{\eta}(x)$ we find to be given by
\begin{eqnarray}
u_{\eta}(x) &=& A_u \left[ \ln \left( {M_u^2 +\Lambda^2 \over M_u^2}
\right) - { \Lambda^2 \over M_u^2 + \Lambda^2} \right. \nonumber \\
&+& \left. \sum_{n=1}^\infty [x(1-x)]^n (n+1)
\left( { \alpha^n \over n} - { \beta^n \over n} -
{ \Lambda^2 \over M_u^2 + \Lambda^2} \beta^n \right) \right] \; ,
\end{eqnarray}
where $\alpha = M^2_{\eta}/M_u^2$, $\beta = M^2_{\eta}/(M_u^2 +\Lambda^2)$
and $A_u$ = 0.09077. The same expression holds for $s_{\eta}(x)$ with
the replacements $M_u \rightarrow M_s$ and $A_u \rightarrow A_s = 0.36307$.
Although this series could be rearranged into a polynomial in $x$,
it is easier to express the moments in terms of
Euler complex Beta functions.  In practice,
30 terms in the expansion are kept, providing a
reasonable $0.1 \%$ accuracy.

\section*{Acknowledgments}

Support from DGES (Spain) Project PB98-1367 and by the Junta
de Andaluc\'\i a is acknowledged. RMD is supported in part
by the U.~S. Department of 
Energy Grant DE--FG02--88ER40448.

% Some useful journal names
\def\NCA{{\em Nuovo Cimento} A }
\def\NIM{\em Nucl. Instrum. Methods }
\def\NIMA{{\em Nucl. Instrum. Methods} A }
\def\NPB{{\em Nucl. Phys.} B }
\def\NPA{{\em Nucl. Phys.} A }
\def\PLB{{\em Phys. Lett.}  B }
\def\PRL{{\em Phys. Rev. Lett.} }
\def\PRD{{\em Phys. Rev.} D }
\def\PRC{{\em Phys. Rev.} C }
\def\ZPC{{\em Z. Phys.} C }
\def\JPA{{\em J. Phys.} A }
\def\EPJC{{\em Eur. Phys. J.} C }

%\section*{References}

\end{document}